\documentclass[10pt,preprint]{aastex}
\usepackage{graphicx}

\begin{document}

\title{\bf M31 and Local Group QSO's}

\author{H. Arp}
\affil{Max-Planck-Institut f\"ur Astrophysik, Karl Schwarzschild-Str.1,
  Postfach 1317, D-85741 Garching, Germany}
 \email{arp@mpa-garching.mpg.de}

\author{D. Carosati}
\affil{Armenzano Astronomical Observatory, 06081 Assisi(PG), Italy}

\begin{abstract}
Previous analyses have shown companion galaxies aligned along the minor
axis of M31. The alignment includes some galaxies of higher redshift
than conventionally accepted for Local Group members. Here we look at
the distribution of all high redshift objects listed in a 10 x 10
deg. area around M31. We find not only galaxies of higher redshift but
also quasars along the minor axis of this brightest Local Group
galaxy, Some are an unusual class of low z, quasar-galaxy.  

Previously observers had noted radio sources aligned along the minor
axis of M31. The ejection directions of quasars from active galaxy
nuclei is also along the minor axis within a cone of about 20
deg. opening angle.  It is shown here that the quasar-like and higher
redshift objects associated with M31 are relatively concentrated along
this axis.
 
 M33 also falls closely along the minor axis of M31 and the famous 3C48
and similar redshift galaxy/quasars are seen along a line coming from
this Local Group companion of M31. What appears to be dusty nebulosity
has also been shown to exist along this extended line in the sky.  
\end{abstract}


\section{INTRODUCTION}
 
Considerable evidence has by now been published showing quasars
and high redshift galaxies associated with active galaxies (Burbidge
et al 2003; Arp 1998a; 2003; Arp and E.M. Burbidge 2005). The question
then arises: Do some galaxies show these associations even though they
are at low levels of activity? To address this question we investigate
M31, which is considered to be a prototype, normal Sb galaxy. Being
the closest large galaxy to us, we can also observe to low luminosity
possible objects belonging to M31.

To our surprise, as soon as we plotted the quasars in the area,
there emerged a striking example of alignment of high and low
redshift objects along the minor axis stretching across the Local
Group from M31.

\section{QSO's around M31}

Fig. 1 shows the result of plotting all quasars (z $\leq$ 2.4) listed
in NED inside a 10 x 10 deg square around M31. The 10 innermost
quasars group closely around M31 and are noticeably aligned along its
minor axis. Then there are 9 quasars about 3 degrees distant which are 
distributed in an arc. Unrelated background objects would not be
expected to form such configurations. In addition the apparent
magnitudes written next to each quasar are noticeably brighter than
quasars normally found in other areas of the sky. 

\subsection{Completeness of the NED Catalog}

In the 44 years since their discovery many quasars have been
measured all over the sky and it is interesting to see now what is known about
those that happen to fall near M31. Table 1 here lists 37 NED quasars
within 5 degrees of M31. It is apparent that many of them have been
discovered more than once, e.g. as radio sources, in Hamburg objective
prism, 2MASX infra red, X-ray surveys and occasionally from color. 
These are measures made over the large region in the direction of the
Local Group center but not necessarily specifically on or near M31.
An exception consists of 5 quasars identified from CFHT slitless spectrum
observations of M31 (Crampton et al. 1997) and one near M31 by van den
Bergh (1966). They are marked UV in Table 1, and excluded from
numerical calculations. The rest of Table 1, as recorded between 1 and 5
degrees, should have no selection bias associated with the position of M31. 

But in fact we see that there is an excess of quasars within about 1
to 4 degrees around M 31. We independently verify this by sampling
84.8 square degrees of control fields outside the 5 degree circle. As an
additional test of association we show that within this 5 deg. circle
the quasar density diminishes away from M31 (Fig.2).
 
Finally, and most importantly here, we show (Figs. 3 and 7) that the quasars
nearest M31 define almost exactly the ejection cones along the minor
axis which were empirically obtained from superposing data from
previous quasars associated with active galaxies (Arp 1998a p.87).

Note: The quasars in Table 1 tend to be bright for their redshift as
expected if there is any physical association with M31. (Most of
the apparent magnitudes are in V wavelengths as in V\'eron and Ceti
V\'eron but we have added four in r (mag.) from USNO-A 1.0 positional
identifications. Homogeneity in magnitude systems is to be wished but
quasars are variable and for current purposes aproximate values are adequate.)

Note also: Table 1 prints out what NED designates as QSO's. In five
cases a G is further noted for type. Upon investigation it is found the
object has been classified once as a QSO and once as a galaxy. The reason
for the QSO designation is usually radiation characteristics or the
bright absolute magnitude the object would have at its conventional
redshift distance. As the remarks following indicate, however, the
absolute magnitude definition of a quasar is not empirical but depends
on an assumption concerning redshift. 

The two lowest redshift objects stand out in Fig.1 because at 
z = .120, .134 they have strikingly low redshifts to be
catalogued as quasars. The reason they are so catalogued is that they
have bright apparent magnitudes. {\it But if they belonged to a more
distant galaxy than M31, they would have fainter apparent magnitudes
and not fit the definition of a quasar as having a luminosity of 
M$\leq$ -23 mag. Hence they would not be listed as quasars.} At their
redshift distance they would be near the upper luminosity for galaxies. 

The preceding comment raises a dificulty in the definition of
quasars. It has been a definition depending on a theory that the
redshift means distance. The definition should preferably be an
empirical criterion of morphology and spectroscopic characteristics
which admits of a continuous transition of properties between quasars
and galaxies.

To emphasize the unusual finding of QSO's between z = .120 and .189 we
note that we get a density of 1.1/sq.deg. close to M31. But in the
control areas off the three corners of Figs. 1 and 3 we get only
.01/sq deg. The mean apparent mag. of those near M31 is 17.4
mag. whereas in a 2dF sample in the -30 deg. strip the average
magnitude at such redshifts is 19.0 mag.

If we wait until complete optical surveys such as 2dF and SDSS are done in
our Local Group area we would expect more quasars but they would  
be fainter and probably predominantly background objects. As of
now the Hamburg objective prism survey covers the Local Group area
($|b| >$ 20, Dec.$>$ 0) to almost 18.0 mag. Note 18.0 and 18.4 mag. in
Table 1. It would seem that the quasars down to about 18.0 mag. were
reasonably complete. It is not useful to refer to quasar densities
in other areas of the sky because in the direction of the Local Group
quasars are very different. For example quasars of z $>$ 1.0 are 6
times their average density in the direction of the Local Super
Cluster (Sulentic 1988a,b). The quasars in the M33 direction are much
different in magnitude and redshift than in the Virgo Cluster
direction (Arp 1984b). In any case it is unlikely that there are a
large number of unmeasured QSO's in the 17 to 18 mag. range which
could erase the concentration around M31.

\begin{figure}[h]
\includegraphics[width=15.0cm]{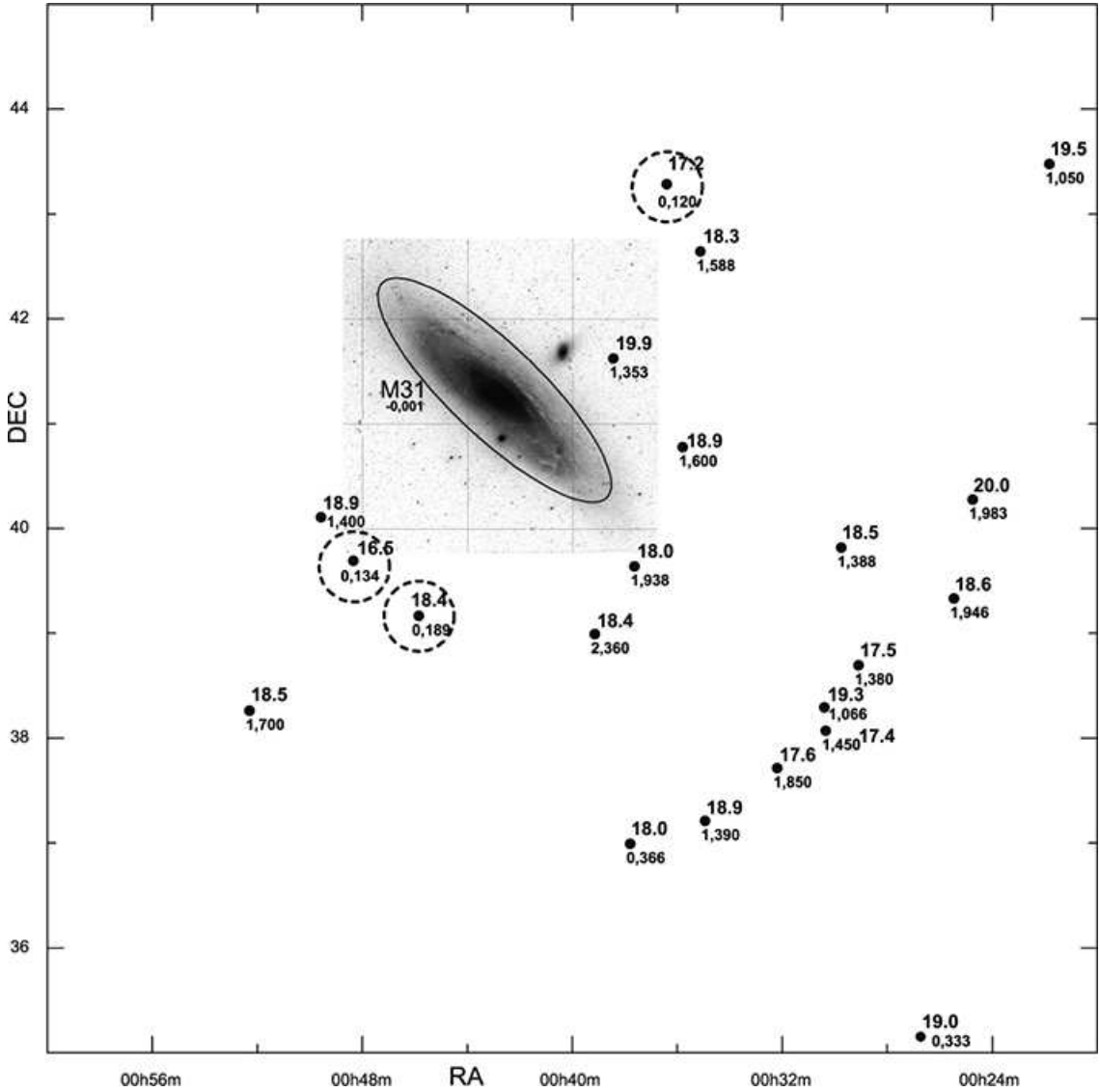}
\caption{Quasars with z $<$ 2.40 in a square 10 x 10 deg. around M31.
\label{fig1}}
\end{figure}

\begin{table}[pt]
\caption{Known QSO's around M31} 
\label{Table1} \vspace{0.3cm}
\begin{tabular}{lcccl}
Object & r(arc min.) & z  & mag. & Remarks\\
& & \\
B3 0037+405   &  38.5 & . . .  & 19.3 & Radio, X-ray, BL Lac\\
B3 0035+413   &  53.0 & 1.353  & 19.9 & Radio, X-ray, var\\
NGC 224 E R   &  64.9 & . . .  & 20.5 & UV\\
MLA93 0034    &  68.9 & . . .  & 17.2 & Obj. Prism, red\\
NGC 224 E J   &  71.4 & 2.400  & 20.2 & UV\\
NGC 224 E A   &  84.3 & 1.600  & 18.9 & UV\\
NGC 224 C29 U & 104.4 & 1.400  & 18.9 & UV\\
HB89 0045+395 & 105.3 &  .252  & 17.4 & Radio, IR, r =16.5\\
HB89 0034+393 & 114.1 & 1.938  & 18.0 & Radio\\
IO And        & 114.3 &  .134  & 16.5 & Vis, IRS, X-ray, var\\
HB89 0032+423 & 118.4 & 1.588  & 18.3 & Radio\\
B3 0050+402B  & 130.2 &  .149  & 18.8 & Radio, IRS, 15.5 I.D.:\\
NGC 224 C29 D & 131.1 & 2.400  & 19.3 & UV\\ 
HB89 0043+388 & 131.2 &  .189  & 18.4 & UV\\
RX J0049+3931 & 132.7 & . . .  & 18.3 & X-ray, HS\\
HS 0033+4300  & 139.7 &  .120  & 17.2 & Obj. Prism\\
HS 0036+3842  & 143.0 & 2.360  & 18.4 & Obj. Prism, IRS\\
NGC 224 C29 B & 143.1 & 2.900  & 20.2 & UV\\
B2 0027+39    & 172.2 & 1.388  & 18.5 & Radio\\
HS 0035+4405  & 193.0 & 2.710  & 17.0 & Obj. Prism, IRS\\
GB6J0052+4402 & 198.3 & 2.623  & 18.3 & Radio\\
4C +37.03     & 211.5 & 1.700  & 18.5 & Radio\\
B2 0022+39A   & 213.3 & 1.983  & 20.0 & Radio\\
HS 0026+3824  & 220.4 & 1.380  & 17.5 & Obj. Prism, IRS\\
HS 0058+4213  & 222.2 &  .190  & 16.0 & Obj. Prism, IRS, blue\\
B2 0027+38    & 228.6 & 1.066  & 19.3 & Radio\\
B2 0022+39B   & 229.7 & 1.946  & 18.6 & Radio\\
HS 0042+3704  & 236.1 & 2.410  & 17.6 & Obj. Prism, IRS\\
B3 0027+377   & 239.8 & 1.450  & 17.4 & Radio, X-ray\\
HS 0029+3725  & 246.2 & 1.850  & 17.6 & Radio, IRS\\
GB6 J0034+3712& 260.0 & 1.390  & 18.9 & Radio\\
87GB 0100+4306& 262.6 & . . .  & 16.6 & Radio, IRS, red\\
4C 36.01      & 263.4 &  .366  & 18   & Radio, X-ray, IRS\\
HS 0058+3820  & 265.4 & 1.920  & 18.0 & Obj. Prism\\
4C +43.01     & 267.1 & 1.050  & 19.4 & Radio, IRS\\
RGB J0042+366 & 275.0 & . . .  & 17.5 & Radio, X-ray, HS\\
87GB 0025+4458& 289.8 & 1.050  & 17.6 & Radio, X-ray\\
 
\end{tabular}
\end{table}   

\newpage

\begin{figure}[h]
\includegraphics[width=14.0cm]{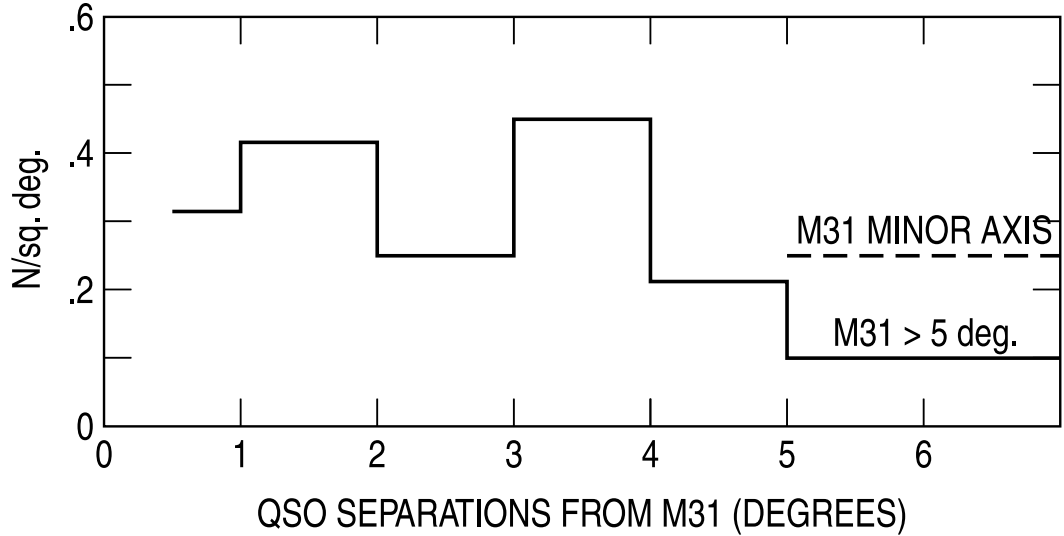}
\caption{Average quasar density (N/sq.deg.) in annuli centered on
M31. Quasar densities diminish going away from M31 but less so along minor
axis alignment. UV selected quasars omitted.
\label{fig2}}
\end{figure}

\subsection{Density of Quasars}

Catalogued quasars of all z values are pictured in Fig. 3. We measure
their density in concentric annuli around M31. As Table 2 shows, the
maximum occurs about N = (.42 - .45)/sq.deg. between 1 and 4 degrees
from the galaxy. By about 5 degrees radius the density has fallen to
around N = .11/sq.deg. We note that this is a minimum over-density
because there are three candidate QSO's at 38.5', 68.9' and 132.7'
whose redshifts have not yet been measured (although control fields
also contain QSO's missing redshifts).  

\begin{figure}[ht]
\includegraphics[width=13.0cm]{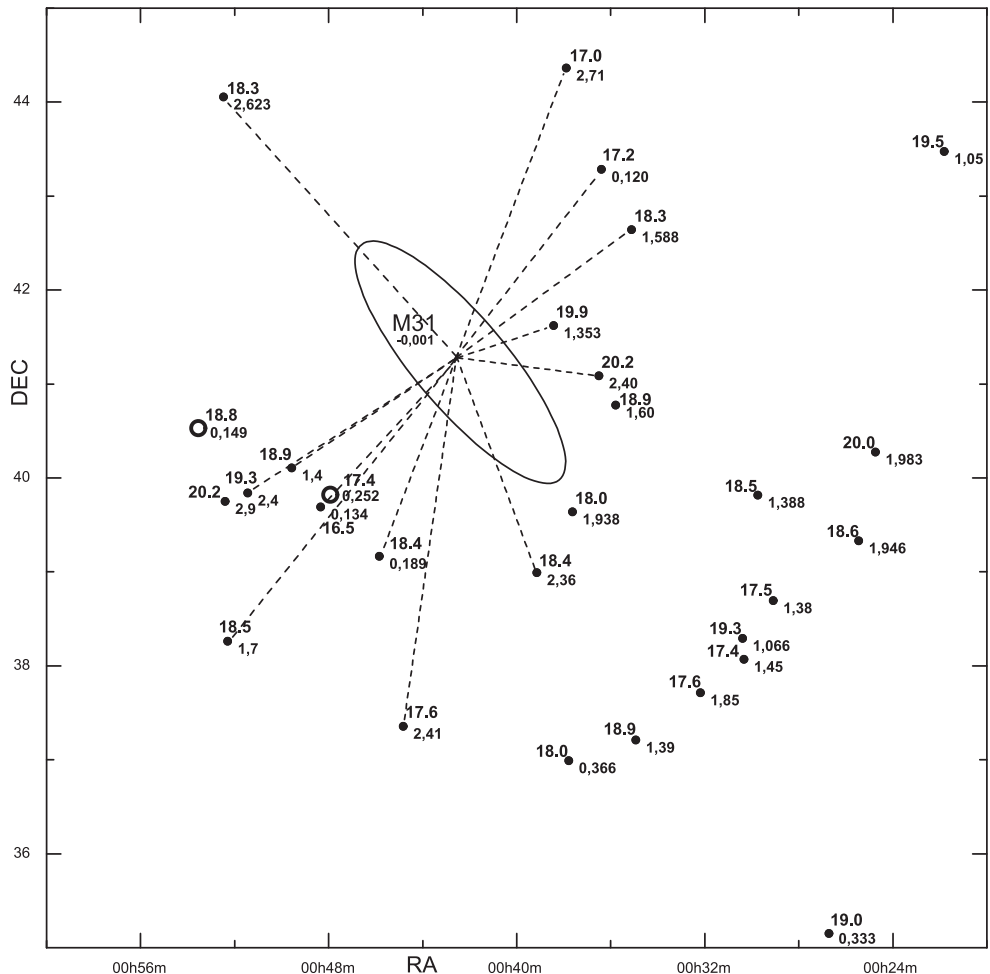}
\caption{All catalogued quasars with suggested pairings indicated by
dashed lines. Redshifts and apparent mags. labeled.
\label{fig3}}
\end{figure}

In the histogram in Figure 2 we would like a sample of background
density even further outside the 5 degree radius we have used for the
extent of M31. We take four circular areas of 3 deg radius
off the four corners of the square in Fig. 3. The R.A. and Dec. of
these 28.27 sq.deg. areas are listed in Table 2.     
                               
The 01h 12m +34d 00m density of N = .32 seemed disturbingly large
until it was realized that sample fell directly in the
extended alignment of M31 companions along the SE minor axis. We find
the quasar density along the full extent of the minor axis 
alignment to be N = .25/sq.deg. This density is 2.3 times the density
off the axis all the way to 3C120. The density for the three fields
away from this axis was (.14 +.18 + 0)/3 = .11/sq.deg.

One way to estimate the significance of the excess quasars around M31
is to calculate the ratio of the numbers of quasars within the 0 to 5
degree radius to the numbers expected from the fields well separated
from M31. Since the effect is maximum in the brighter quasars we   
take quasars with magnitudes $<$19 and omit 3 UV quasars in Table
1 (equaling 21). We then average the three control fields from Table 2,
only for mag.$<$19 which yields (.14 + .11 + 0) = .083/sq.deg. Then the
significance for the excess around M31 is:

                $$(21 - 6.52)/\sqrt 6.52 = 5.7 sigma$$ 

But if we take all QSO's, bright and faint, without UV QSO's, we
get a background density of .11/sq.deg. and a 5.6 sigma result. So
this is not so sensitive to the magnitude limit adopted. Moreover as
we note in Section 2.5 absorption in M31 raises the significance of the
excess. We only emphasize here that we find a significant over density
of quasars between 1 and 4 deg. from M31 as is visually apparent from
the absence of quasars toward the outer edges of Figs. 1 and 3. Note
also the apparent ring of 6 quasars 2.4 $\leq$ z $\leq$ 2.7 around M31
in Fig.3. The low z glaxies also show association with M31, however,
and since they are are unusual it is interesting to investigate them
in more detail.

\begin{table}[ht]
\caption{Quasar Density Around M31} 
\label{Table2} \vspace{0.3cm}
\begin{tabular}{lcccl}
Annulus & Area(sq.deg.) & N (QSOs) & N/sq.deg. & Remarks\\
& & \\
0 to 1 deg. & 3.142 & 1 & .32 & obstructed by M31\\
1 to 2 deg. & 9.425 & 4 & .42 & \\
2 to 3 deg. & 15.71 & 4 & .25 & \\
3 to 4 deg. & 21.99 &10 & .45 & \\
4 to 5 deg. & 28.27 & 6 & .21 & \\
r = 3 deg.\\
00h08m+34d00m & 28.27 & 4 & .14 & off corners of M31 square\\
00h08m+46d00m & 28.27 & 5 & .18 &       ``            ``   \\
01h12m+46d00m & 28.27 & 0 & .00 &       ``            ``   \\
01h12m+34d00m & 28.27 & 9 & .32 & M31 minor axis alignment \\
NGC 404       & 28.27 & 7 & .25 &       ``          ``     \\
NGC 918       & 28.27 & 7 & .25 &       ``          ``     \\
3C120         & 28.27 & 7 & .25 &       ``          ``     \\ 
\end{tabular}
\end{table}

\newpage

\subsection{Note on low z, quasar-like objects}

It has already been noted that some of the objects near M31 have
unusually low redshifts and are only classified as quasars because of 
their bright apparent magnitudes. If they were slightly fainter they
would be calculated to be fainter than the conventional quasar limit of
-23 mag. In fact there are other objects near M31 which fall slightly
below this arbitrary cut off but which have spectra and morphology very
much like quasars. Two examples are shown as small open circles in Fig. 3.
One has z = .149 and  a listed magnitude of 18.8 mag. But its DSS 
magnitudes are r = 15.5, b = 17.3 mag. On a luminosity definition it
could, paradoxically, be a QSO in the red and galaxy in the blue. It is
listed as strong emission (Djorgovski et al. 1995) and there are
extremely straight radio ejections coming from the central
object (B3 0050+401). The other is B3 0045+395 at z = .252. At
apparent mags. r = 16.5 and b = 17.4 mag., and a designation BL Lac?, it
certainly is in the conventional quasar category although classed as
Sey 1. With these two added to the previously mentioned very low
redshift quasars one can see that there is an indication of an unusual
kind of quasar associated with M31, the kind of an object that would not
be classed as a quasar in a more distant galaxy. On the other hand,
the z = .189 object, though usually called a quasar, fails the absolute
magnitude criterion. 

In the 3 deg. field around NGC 404 (also along the M31 minor axis to
M33) there is a z = .107 object
at 15.5 mag. (2MASX J01174564+3637145) which is labeled QSO in
VCV 2001 (V\'eron-Cetti and V\'eron 2001). Actually these kind of low
redshift quasars are noticeable around large nearby galaxies. Two at z
= .215 and .216 are shown to be associated with the active, ejecting
NGC 3079 (Arp and Burbidge 2005). It could be a worthwhile
exercise to study just this interval around nearby galaxies versus
more distant galaxies where the objects belonging to the galaxy drop
in apparent magnitude below the quasar definition.

\subsection{Low Luminosity Quasars}

If some of the low redshift objects near M31 fall just
marginally brighter than the conventional quasar criterion of
luminosity, then the question arises: What about objects that are
just sightly fainter than this level? Are there more galaxies listed
near M31 which resemble quasars in their radiative characteristics?
Surprisingly, Fig. 4 shows there are indeed many galaxies, .05 $<$ z
$<$ .25, which fall close to the M = -23 mag. line. It suggests that
there would be more quasars listed around M31 if their definition did
not depend on an arbitrary luminosity which in turn is based upon an 
inapplicable redshift/distance criterion. 

\begin{figure}[h]
\includegraphics[width=12.0cm]{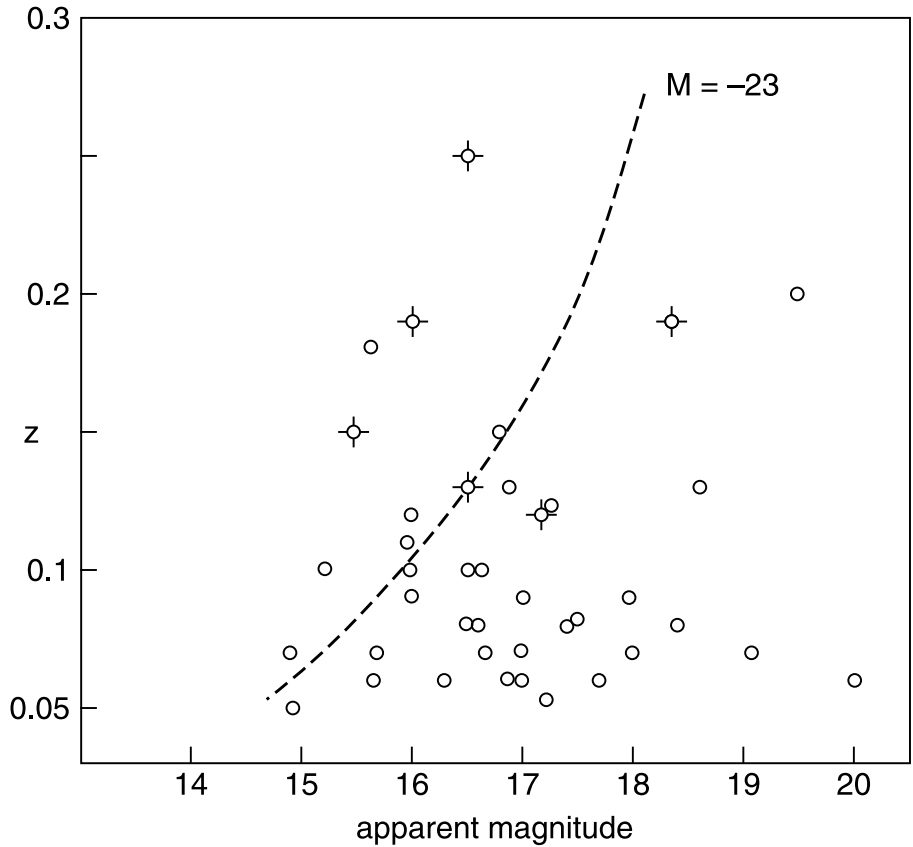}
\caption{z/m for galaxies $\leq$ 5 deg. of M 31. NED redshifts $.05 < z <
.26$. Low z quasar properties marked with plus signs. 
\label{fig4}}
\end{figure}

{\bf Suggested New Definition of QSO:} {\it A quasar is a compact (high
surface brightness) object which shows appreciable non
thermal radiation.} This would make the spectroscopic or multi wavelength
classification be entirely empirical and enable one to study the
quasar/galaxy continuity over a range in luminosities. This is already
being done for very high redshift galaxies which have just as
extraordinarily high redshift as some of the highest redshift quasars
but show an underlying stellar spectrum.

Independently of their possible quasar-like properties, however, the 
important result shown in Fig.5 is that 40 objects classified as
{\it galaxies} by NED are physically associated with M31 despite their mean
redshift being z $\sim$ .1. They are even more strongly
concentrated around M31 than the quasars in Figs. 1 through 3.

\begin{figure}[h]
\includegraphics[width=10.0cm]{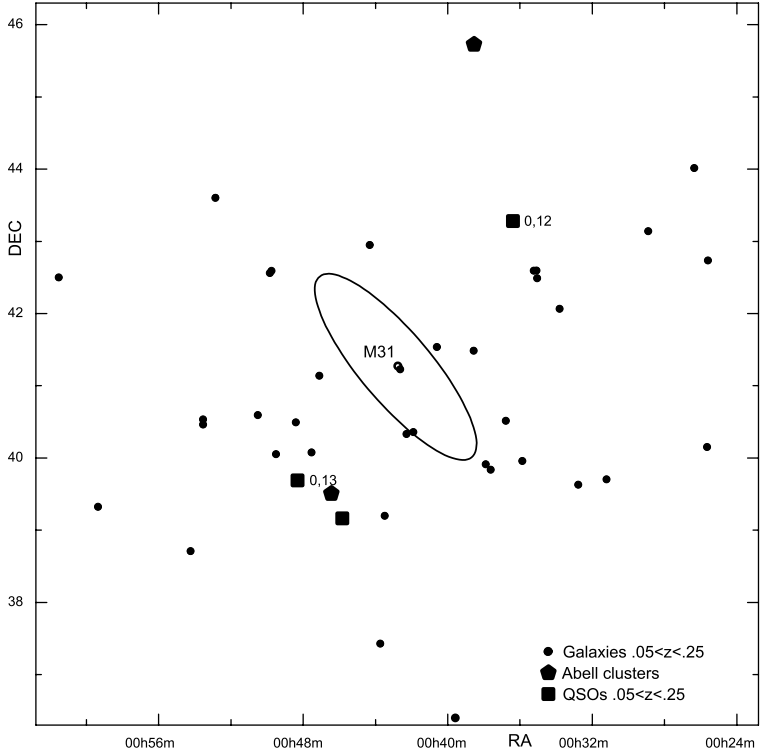}
\caption{All galaxies with NED redshifts $.05 < z < .26$. Low z quasar
pair labeled z = .12 and .13.
\label{fig5}}
\end{figure}

Question: How do we know the concentration of galaxies around M31 in
          Figs.5 and 6 is not a background cluster of galaxies?

Answer: These 40 galaxies are spread out within a diameter of about 8
        degrees on the sky. They have an apparent diameter of the
        order of the Virgo Cluster but a mean redshift about 30 times
        greater! As a single cluster it would be too large and the
        internal range in redshifts would be too large. As for
        multiple clusters in just this line of site there is no
        evidence for groups of redshifts at discrete values. Finally
        there is their strongly increasing density toward the center
        of M31. 

It is informative to note that 22 outof the 40 galaxies plotted in
Fig.5 are infrared sources from the all sky, 2MASX survey. Then there
are 5 radio galaxies, 4 MCG, 2 Zwicky, 1 Mrk etc. In addition to the
size of the association, signs of activity in these galaxies around
M31 are already becoming evident.

Fig.6 shows the increase in galaxies per sq.deg. as annuli closer to
the center of M31 are measured. Table 3 gives the numerical
densities in the rings.  
  
\begin{table}[ht]
\caption{Density of .05 $<$ z $<$ .26 Galaxies Around M31} 
\label{Table3} \vspace{0.3cm}
\begin{tabular}{lcccl}
Annulus & Area(sq.deg.) & N (QSOs) & N/sq.deg. & Remarks\\
& & \\
0 to 1 deg. & 3.142 & 6 &1.91 & obstructed by M31\\
1 to 2 deg. & 9.425 &15 &1.59 & \\
2 to 3 deg. & 15.71 & 8 & .51 & \\
3 to 4 deg. & 21.99 & 7 & .32 & \\
4 to 5 deg. & 28.27 & 4 & .14 & \\
\end{tabular}
\end{table}

If we take the outermost ring reading as the density of galaxies
unrelated to M31, then we can compute the chance of getting 40 within
a 5 degree radius (78.54 degree area) from M31. The calculation gives:

        $$ (40 - 11)/\sqrt 11 = 8.7 sigma$$

Even this extremely large sigma is an underestimate of the significance
of the association because it does not take into account the
monotonic increase in density as the annuli approach M31. Neither does
it allow for the obscuration of the background from dusty absorption
from M31. The latter could be considerable considering the apparent
size of M31.

\subsection{Absorption from M31}

The disk of M31, in which the dusty absorption is concentrated, is
tilted within about 9 deg. of edge on to the observer. One can see
from the image of M31 (e.g. Fig. 1) that the near side of the disk is
on the NW edge of the ellipse as we see it. This means that the minor
axis to the NW is more obscured than the minor axis to the SE. Is this
reflected in the number and apparent magnitude of the objects shown
in Figs. 3 and 5? Yes. Approximately 5 objects NW in Fig. 3 have
average mag. 18.9 while 8 along the SE minor axis have 18.4 mag. The
imbalance of galaxies NW re SE is also vsually apparent in Fig.5.

\begin{figure}[h]
\includegraphics[width=10.0cm]{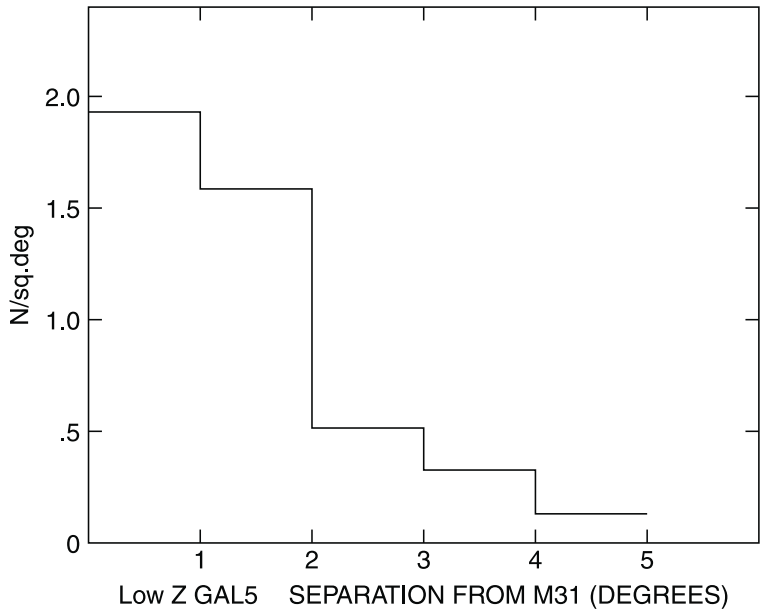}
\caption{All galaxies with NED redshifts $.05 < z < .26$. Density in
N/sq.deg. in annuli centered on M31.
\label{fig6}}
\end{figure}

If the cloud of galaxies under discussion were at its redshift
distance it would contain of the order of a dozen galaxies in the
$M_{abs}$ = -23 mag. range (See Fig.4). This is quite unprecedented
since the brightest central galaxies we know are in the -21, -22 mag. 
range. On the other hand, at the distance of M31 the galaxies would be extreme
dwarfs in the -7 tp -9 mag. range. (For systematic evidence on the
association of clusters of galaxies with bright, nearby galaxies see
Arp and Russell 2001). One suggestion would be that they may
represent quasars (which are underluminous to start with if associated
with nearby galaxies) that have broken up in the ejection from the
interior of M31 and are evolving to, perhaps, globular clusters as
they approach the age of the Local Group galaxies. At z = .1, a late
stage in their evolution, these still somewhat active objects have no
where to evolve except possibly to clusters which have a range of
absolute magnitude with recent star forming medium having been stripped
out by various encounters with clouds along the line of exit from
the nucleus of the parent galaxy.

\begin{figure}[h]
\includegraphics[width=13.0cm]{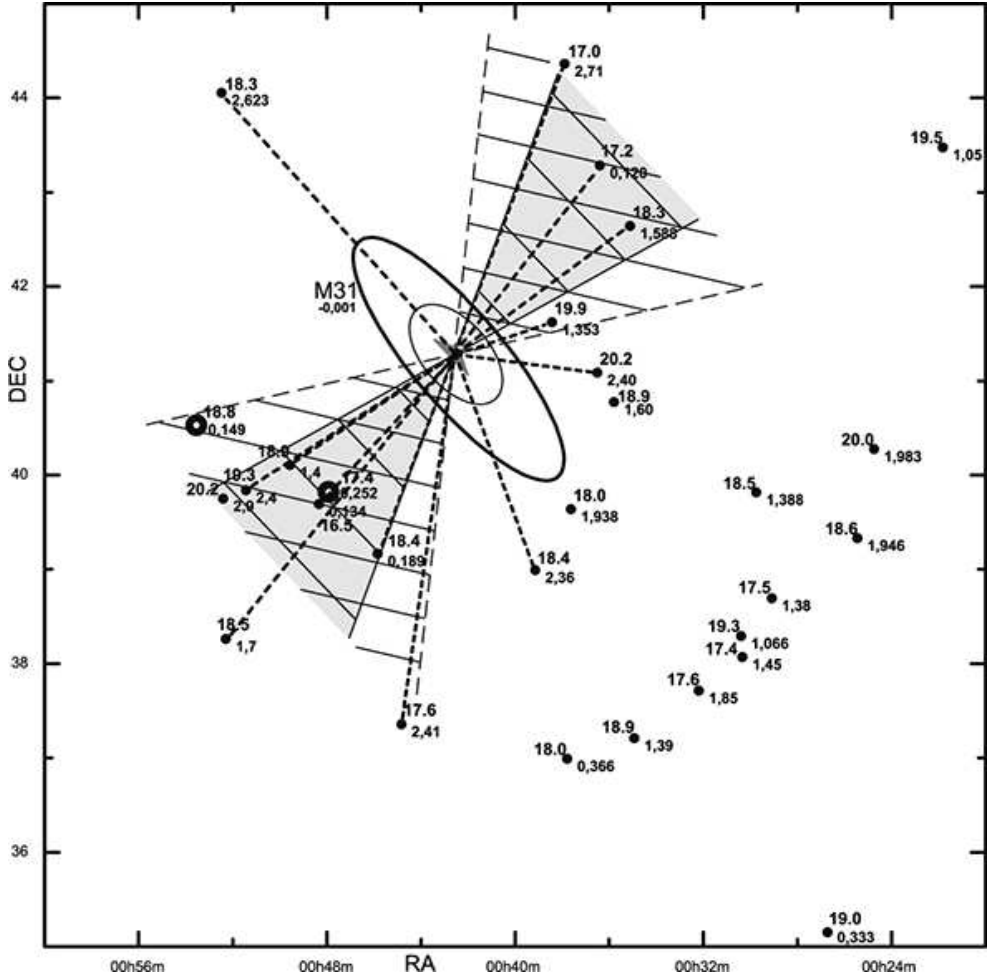}
\caption{All quasars from Fig.3 with superposed ejection patterns from
previous active galaxies.
\label{fig7}}
\end{figure}

\subsection{Quasars along the minor axis}

In addition to the over density of quasars and near quasars around M31
there is also the point that they are elongated in their distribution
and that elongation is markedly along the well defined minor axis of
M31. Fig. 7 shows the empirically defined ejection cones from combined
data on active galaxies.

The half opening of the cone angle for quasars ejected from active
galaxies was observed initially as $\pm 20$ deg. with $\pm 35$ deg. for
companion galaxies (Arp 1998a, p87). Most recently L\'opez Corredoira and
Guti\'errez (2007) have confirmed a similar angle of emergence along the
minor axis from large survey samples of quasars near galaxies. It is
striking that the M31 data here further support these previous
results. It is also important for models of evolution from high
redshift quasars to low redshift galaxies. Note also the evidence for
ejection of radio material along the M 31 minor axis in section 4 of
this paper.

\subsection{Pairing of quasars across M31}

While Fig. 1 omitted quasars with z $\geq$ 2.40, Fig. 3 shows all
QSO's including the highest redshifts. Lines connecting apparent pairs
have been dashed and emphasize the equal and opposite ejection
events which are hypothesized to account for this pervasive property
of quasars associated with galaxies.(Arp 2003).

\begin{table}[ht]
\caption{Quasar Pairs across M31} 
\label{Table4} \vspace{0.3cm}
\begin{tabular}{cccl}
z & mag. & z(ave) & Remarks\\
& & \\
.120 & 17.2 & .127 & close alignment\\
.134 & 16.5 & &          \\
                         \\
2.623 & 18.3 & 2.49 & $z_K$ = 2.64 \\
2.360 & 18.4 & & \\
                         \\
2.710 & 17.0 & 2.56 & $z_K$ = 2.64 \\
2.410 & 17.6 & \\
                          \\
1.588 & 18.3 & 1.644 \\
1.700 & 18.5 &       \\
                           \\
1.600 & 18.9 & 1.500 & $z_K$ = 1.41 \\
1.400 & 18.9 & \\
\end{tabular}
\end{table}

It is impressive to note the similarity of the redshifts and apparent
magnitudes of the quasars in the pairs. Some are listed in Table 4.
They also tend to be brighter in apparent magnitude than average quasars 
observed over the sky. (Note the pair of z = 2.710 and 2.410 at the
very bright 17.0 and 17.6 mag. There is also a tendency for the
average redshift in the pair to fall close to two of the Karlsson
preferred redshift peaks at 1.41 and 2.64 (Arp et al. 1990). 

Perhaps of greater importance it is clear that the pairs of quasars
align within a cone extending in both directions along the minor
axis. We will see in the following sections that the accepted
companions, higher redshift companions and quasars, extend in the SE
minor axis direction far across the across the sky in this Local Group 
direction.

\newpage
                       
\section{ALIGNMENT ALONG M31 MINOR AXIS}


Extensive early studies of edge-on disk galaxies showed concentrations
of companions along their minor axes (Holmberg 1969, p.317). Later a
very conspicuous alignment of smaller galaxies along the minor axis of
M31 was reported (Arp 1990, p.434, 1998b, p.662). As Fig. 8 here shows,
the alignment consists of all the brightest conventional members of the
Local Group.

\begin{figure}[h]
\includegraphics[width=11.0cm]{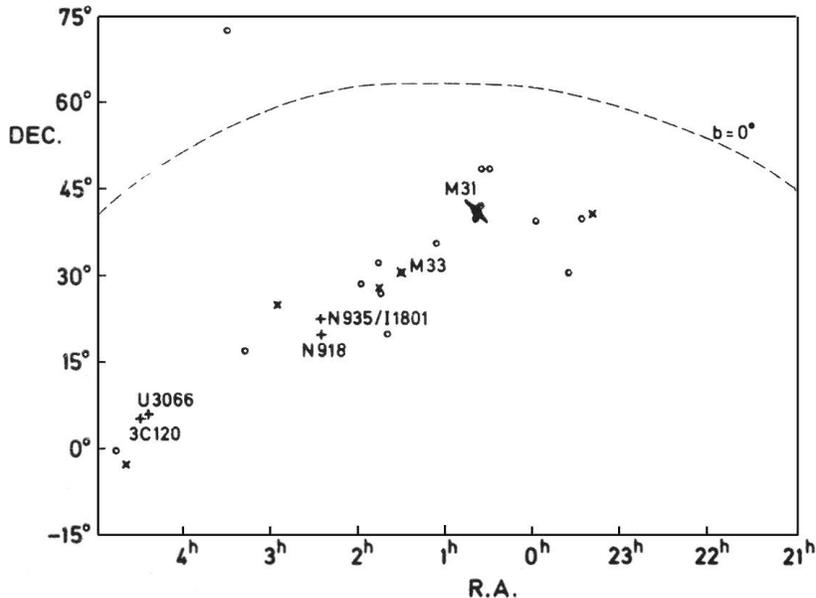}
\caption{Conventional members of the Local Group (-86 $\leq v_0 \leq$ +140
km/sec) are filled symbols. All galaxies known with 300 $\leq v_0 \leq$
700 km/sec are shown by crosses for spiral, open symbols for dwarfs. Some
higher redshift, peculiar galaxies are plus signs SE of M33. See Arp 1990.
\label{fig8}}
\end{figure}                   

If we take galaxies with slightly higher redshifts than acccepted for
the Local Group, say 300 $\leq$ cz $\leq$ 700 km/sec, we see in Fig. 8
that they all fall along the same minor axis alignment from M31. This
establishes some important points: 

1) Bright companions to M31 are almost exclusively distributed along
a narrow line along the SE minor axis. 

2) These companion galaxies have distinctively small redshifts but are
systematically redshifted with respect to M31 (Arp 1994, 1998b).

3) 12 galaxies SE of M33 extend back to M31 accurately along the
line  of the M31 minor axis. These galaxies have redshifts of $v_0$ =
647, 1625 and the rest 4116 through 5163 km/sec. They are almost all
of peculiar, non equilibrium morphology. (Arp 1990, p. 435)

4) The terminus of the line from M31 is 3C120 and UGC 3066 (v = 9,896
and 4640 km/sec). This terminus  is accurately at p.a. 125 deg.  - the
same as the minor axis of M31. 3C120 is a much studied, quasar-like,
strong radio source.
    
In general the distribution of companion galaxies around
minor axes tended to be, as Holmberg had pointed out, about $\pm$35
deg. This is the opening angle of the  ejection cone defined by
the objects in Figs. 3 and 7. But why then is the alignment
of galaxies so narrow over a longer extent in the Local Group?  

We comment that if a high intrinsic redshift (Narlikar and Arp 1993)
proto quasar comes out exactly along the rotation axis it can
travel a long distance while it is evolving to more mass and lower
intrinsic redshift i.e. becoming a well aligned galaxy. If the
ejection is somewhat off the exact rotation axis of the ejecting
nucleus then it encounters medium and clouds which slow, deviate and
break up the proto quasar/galaxy plasmoid. These products are small
and stay close to the ejecting galaxy as in the picture of M31 in Fig. 7.
But it also implies that the ejection axis in M31 has operated
over a period of time comparable to the evolution from compact proto
quasars to normal, only slightly redshifted companion galaxies.

\section{RADIO SOURCES ALONG THE M31 MINOR AXIS}

Early searches for a radio halo around M31 revealed instead point
sources concentrated along the M31 minor axis. To quote Wielebinski
(1976): ``By a strange chance of nature there are groupings of rather
strong sources in nearly symmetrical positions on opposite sides of the
nucleus along the minor axis.'' (See also Wielebinski 2000). The
importance of this finding lies in the accepted origin of radio
sources as ejections from centers of galaxies. Since we have seen
companion galaxies along this M31 minor axis as well as higher
redshift galaxies and quasars it is reasonable to conclude this is an
ejection path for material which evolves into normal companion galaxies. 

Confirmation of this ejection of radio sources can be seen in the
408 MHz, 3.9 x 2.7 deg. map of M31 (Gr\"ave 1981). In their Fig. 3
just laying a straight edge at p.a. 125 deg. shows that essentially
all of the sources lie along the M31 minor axis.

\section{NEBULOUS DUST ASSOCIATED WITH M31 EJECTION}


Finally we come to the aspect which could most shake conventional
beliefs about the Local Group and the nature of near space. Deep
prints of a red sensitive schmidt plate (Arp and Sulentic 1991) show  
unmistakable filamentary dust features reaching back along the minor
axis direction toward M31.This filament is repeated in the blue
photographs and 100 the hundred micron infra red scans. They have to
be real. Although no one has cared to take a spectrum there is no hint
of gaseous emission.

The ejection path across the whole Local Group sky from M31 to 3C120
(Note the well known but little attended nebulosity just north of
3C120) must have carried material
either dusty or capable of forming dust from the ejecting M 31. But
that means dust and obscuration within the Local group of galaxies - a
point which has never before been seriously advanced. But how can one
escape the multiwavelength evidence and the detailed discussion
of Arp and Sulentic (1991)?

The most provocative object in the M31 minor axis line is NGC 918 at $v_0$
= 1640 km/sec redshift. The nebulous dust is most concentrated at the
position of the galaxy but a region has been cleared on either side of
the minor axis of the galaxy. Higher resolution images would give
invaluable information on the process whereby ejections come out along
the minor axis of galaxies.

In addition the nebulosity is of such long extent across the sky and
so coincident with the alignment along the M31 minor axis that it must
be in the Local Group. Therefore interaction with the dust filament
would represent direct evidence for a distance much smaller than NGC
918's conventional redshift distance.

Fig. 9 shows a 29 x 29' region around NGC 918. It is taken with the
40cm reflecting telescope of D. Carosati at the Armenzano Observatory in
Assisi, Italy, with 50 summed exposures in the R band totaling 150
minutes.  

The filamentary features surrounding NGC 918 are well shown in this
image. The outer features appear to be dust illuminated by the galaxy.
Immediately around the galaxy the dust appears to cleared away. By
either outflow of matter or radiation pressure from the galaxy.

If further imaging confirms that the galaxy is actually interacting
with the faint surface brightness material then we have a galaxy of cz 
= 1640 km/sec redshift at the same distance of nebulous matter as the
Local Group of Galaxies which is essentially cz $\sim$ 0 km/sec.

If the galaxy is not interacting with the nebulosity but just shining
through a serendipitous hole we still have the remarkable inference
that material has been ejected along the minor axis of M31 into the
middle of the Local Group of galaxies. The question then arises as to
how many other nearby galaxy groups contain intergalactic material and
what this would do to our view of purportedly more distant galaxies.

\begin{figure}[h]
\includegraphics[width=10.0cm]{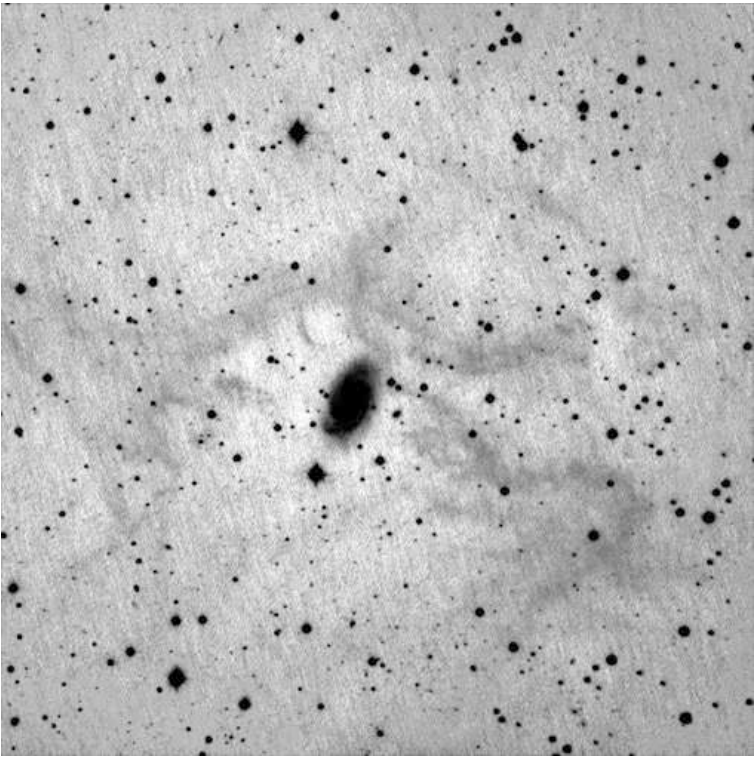}
\caption{Faint surface brightness features around NGC918 (29 x 29').
\label{fig8}}
\end{figure}

\section{Remarks on NGC 404 and M33}

NGC 404 falls approximately midway between M31 and M33 and almost
exactly on the minor axis line back to M31. It is an unusual galaxy in
that it has a presumably old halo of red stars but at the center has a
compact, high surface brightness nucleus. Its redshift is +228 km/sec
greater than M31.


M33 is well known Local Group spiral galaxy. It also lies on the
alignment of companions back to the nucleus of M31. It is considered a
normal, prototype spiral galaxy and, as in M31, it is of interest to
see if there are any associations with quasar-like
objects. Considering all the known quasars within a region of about 7
x 10 degrees it is seen that there is a fairly clear region around M33
as in many active parents and a pair of very high redshift quasars on
either side
of the galaxy. (z = 4.532 nad 4.258) Such high redshifts are rare and
it is noticed that their apparent magnitudes, like the quasars around
M31, are quite bright. If these two had intrinsic redshifts at the
Karlsson peak of z = 4.51 then their velocities of ejection would be
$z_v$ = + .005 and -.046. We note, however, a quasar with z = 4.220
further from M33 to the NW.

{\bf 3C48 and M33}

The most striking aspect of the neighborhood of M33 is that one of
the closest quasars to M33 is the famous 3C48, in 1963 the first
discovered quasar. See the three investigations in 1984 (Arp
1984a,b,c) of the Local Group direction in the sky. There it was shown that
M33 was at the  head of a line of radio quasars and galaxies of similar
redshift (0.27 $\leq$ z $\leq$ 0.47) and apparent magnitude. The above
three papers showed that this region in the SGH had extensive radio
mapping and that a number of radio quasars had been identified. In
fact, because of its nebulous image and strong radio emission the
prototype quasar, 3C48, acted more like a member of a group of radio
galaxies (albeit with large range in redshifts). 

Ironically 3C48, the much publicized quasar that astronomers fiercely
competed for in 1963, turns out to be just a compact radio galaxy
(with an angular extent of 12'') but with a redshift which conventionally
would place it at great distance in back of M33. The redshift -
magnitude plot for quasars in the Local Group vs those in the
Local Supercluster direction should be studied, however, because they
are very different (Arp 1984a).

Of course 3C48 was first discovered because of its strong radio
emission and bright apparent magnitude. If quasars are associated with
galaxies it would seem likely that 3C48 would be associated with one of the
brightest galaxies in the sky. In the preceding paper we have now given
evidence for quasars being emitted from M31, the central galaxy in the
Local Group. It would seem supportive of that finding to find the
second brightest galaxy in our Local Group also associated with higher
redshift quasars. 

If M33 were removed to distances of
fainter galaxies its apparent separation between it and 3C48 would be
smaller and its apparent magnitude would be fainter, about where we
begin to observe such objects asssociated with more distant galaxies.


\newpage

\section{Summary}

The evidence for association of higher redshift galaxies and quasars
with M31 and the Local Group is:

1) Excess of quasars of bright apparent magnitude within 5 deg. radius of M31.

2) Increasing density of quasars from 5 to 1 degrees toward M31.

3) Unusual class of low redshift quasars.

4) Distribution of associated quasars in two cones along M31 minor axis.

5) Pairing of redshifts and magnitudes across nucleus.

6) Ejection of radio material and alignment of companion galaxies
   along M31 minor axis.

7) M33 located along minor axis of M31. M33 associated with line of quasars
   that are similar to the first-discovered 3C48.

Finally ejection along the minor axis of M31 appears to reach into the
middle of the Local Group and may be involved with medium redshift
galaxies and/or nebulosities.

\section{Conclusions and Criticism}

Excluding gravitational lensing on the basis of the wide separations,
and properties listed above, there are only two possibilities in the
region around M31. Either the catalogued quasars are physically
associated with the galaxy and its aligned companions or there
has been a special search around the galaxy compared to adjacent 
regions. In the latter case it would also be necessary to demonstrate
a systematic bias to sample along minor axis directions or adjacent
arcs. It would be necessary to find observers who said ``I searched for
high redshift objects by looking in the extended neighborhood of large,
low redshift galaxies!

A quite opposite selection effect is more likely.
Because of the almost universal belief in red shift distances local
objects would have been avoided if anything. Criticisms about
``completeness'' or magnitude scales should not be allowed to postpone
further checks with control fields such as have been started in the
present paper.
  
In view of the importance for astronomy and physics it would
seem that selection effects would have to be demonstrated otherwise
the present observational results would need to be seriously
considered and further investigated along with their consequences for
the nature of redshifts.

{\bf Acknowledgement}: In 1965 Allan Sandage published a paper in part
titled ``A Major New Constituent of the Universe: The Quasi-Stellar
Galaxies''. In the ensuing years the QSO's or quasars as they came to
be known, truly developed into a constituent that embraced galaxies in many
evolutionary stages. Still in controversy, the observations discussed 
here perhaps extend the importance and inclusive nature of the
objects he and others at that time reported.

\section{REFERENCES}

Arp, H. 1984a, ApJ 277, L27

Arp, H. 1984b, P.A.S.P. 96, 148

Arp, H. 1984c, J. Astrophys. Astr. 5, 31 

Arp, H. 1990, Astr. Astrophys. 229, 93

Arp, H. 1994, ApJ, 430, 74

Arp, H. 1998a, Seeing Red, Apeiron, Montreal

Arp, H. 1998b, ApJ 496, 661

Arp, H. 2003, Catalogue of Discordant Redshift Associations, Apeiron, Montreal

Arp, H., Bi, H., Chu, Y., Zhu, X. 1990 A\&A 239, 33

Arp, H. and Sulentic, J. 1991, Ap \& Space Sci. 185, 249

Arp, H. and Burbidge, E.M. 2005, astro-ph 0504237

Arp, H. and Russell, D. 2001, ApJ 549, 802

van den Bergh, S. 1966, ApJ 144, 866

Burbidge, E. M., Junkkarinen, V., Koski, A., Smith, H., Hoag, A. 1980,
ApJ, 242L, 55

Burbidge, E. M., Burbidge, G., Arp, H., Zibetti, S. 2003, ApJ 591, 690

Crampton, D., Gussie, G., Cowley, A., Schmidtke, P. 1997, AJ 114, 114,
2353

Djorgovski, S.,Thompson, D.,Maxfield, L., Vigotti, M., Grueff,
G. 1995, ApJS, 101, 255

Gr\"ave R., Emerson, D. T. and Wielebinski, R. 1981, A\&A 98, 270

Holmberg, E. 1969, Arkiv f/''or Astronomii, Vol. 5, p.305-343

L\'opez-Corredoira, M., Guti\'errez, C. M. 2007, A\&A, 461, 59     

Mitchell, K., Warnock A., Usher, P. 1984, ApJ 287, L3

Narlikar, J. and Arp, H. 1993, ApJ 405, 51 

Sandage, A. 1965, ApJ 141, 1560

V\'eron-Cetty, M. and V\'eron, P. 2001, ESO Scientific Report

Wielebinski, R. 1976, A\&A 48, 155

Wielebinski, R. 2000, Proceedings 232 WE-Heraeus-Seminar,
Eds. Berkhuijsen, Beck and Walterbus, p 168

\enddocument